\documentclass[amsmath,amssymb,superscriptaddress,longbibliography,nobalancelastpage,aps,twocolumn,showpacs]{revtex4-1}
\usepackage[normalem]{ulem}
\usepackage{color}
\usepackage{hyperref}
\hypersetup{colorlinks,linkcolor=blue,urlcolor=blue,citecolor=blue}

\usepackage{graphicx}

\newcommand{\LSCO}{La$_{2-x}$Sr$_{x}$CuO$_4$}
\newcommand{\LBCO}{La$_{2-x}$Ba$_{x}$CuO$_4$}
\newcommand{\af}[1]{\textcolor{black}{#1}}

\newcommand{\mh}[1]{\textcolor{black}{#1}}
\newcommand{\mhh}[1]{\textcolor{red}{#1}}

\newcommand{\sakai}[1]{\textcolor{cyan}{#1}}

\makeatletter 
\renewcommand{\fnum@figure}{\textbf{Figure~\thefigure}}
\makeatother

\begin{document}

	
	\title{Pseudogap in electron-doped cuprates: Strong correlation leading to band splitting}

	\author{M.~Horio}
		\email{mhorio@issp.u-tokyo.ac.jp}
		\affiliation{Department of Physics, University of Tokyo, Bunkyo-ku, Tokyo 113-0033, Japan}
		\affiliation{Institute for Solid State Physics, The University of Tokyo, Kashiwa, Chiba 277-8581, Japan}
	
	\author{S.~Sakai}
		\email{shiro.sakai@riken.jp}
		\affiliation{Center for Emergent Matter Science, RIKEN, Wako, Saitama 351-0198, Japan}
		
	\author{H.~Suzuki}
		\affiliation{Department of Physics, University of Tokyo, Bunkyo-ku, Tokyo 113-0033, Japan}
		\affiliation{Frontier Research Institute for Interdisciplinary Sciences, Tohoku University, Sendai 980-8578, Japan}
	
	\author{Y.~Nonaka}
		\affiliation{Department of Physics, University of Tokyo, Bunkyo-ku, Tokyo 113-0033, Japan}
		
	\author{M.~Hashimoto}
		\affiliation{Stanford Synchrotron Radiation Lightsource, SLAC National Accelerator Laboratory, Menlo Park, California 94305, USA}
		
	\author{D.~Lu}
		\affiliation{Stanford Synchrotron Radiation Lightsource, SLAC National Accelerator Laboratory, Menlo Park, California 94305, USA}
	
	\author{Z-.X.~Shen}
		\affiliation{Stanford Synchrotron Radiation Lightsource, SLAC National Accelerator Laboratory, Menlo Park, California 94305, USA}
	
	\author{T.~Ohgi}
		\affiliation{Department of Applied Physics, Tohoku University, Sendai 980-8579, Japan}
	
	\author{T.~Konno}
		\affiliation{Department of Applied Physics, Tohoku University, Sendai 980-8579, Japan}
	
	\author{T.~Adachi}
		\affiliation{Department of Engineering and Applied Sciences, Sophia University, Tokyo 102-8554, Japan}
	
	\author{Y.~Koike}
		\affiliation{Department of Applied Physics, Tohoku University, Sendai 980-8579, Japan}
	
	\author{M.~Imada}
		\email{imada@g.ecc.u-tokyo.ac.jp}
		\affiliation{Department of Engineering and Applied Sciences, Sophia University, Tokyo 102-8554, Japan}
		\affiliation{Research Institute for Science and Engineering, Waseda University, Shinju-ku, Tokyo 169-8555, Japan}
		\affiliation{Toyota Physical and Chemical Research Institute, Nagakute, Aichi 480-1192, Japan}
	
	\author{A.~Fujimori}
		\email{fujimori@phys.s.u-tokyo.ac.jp}
		\affiliation{Department of Physics, University of Tokyo, Bunkyo-ku, Tokyo 113-0033, Japan}
		\affiliation{Center for Quantum Science and Technology and Department of Physics, National Tsing Hua University, Hsinchu 30013, Taiwan}

	
	\begin{abstract}
		The pseudogap phenomena have been a long-standing mystery of the cuprate high-temperature superconductors. The pseudogap in the electron-doped cuprates has been attributed to band folding due to antiferromagnetic (AFM) long-range order or short-range correlation. We performed an angle-resolved photoemission spectroscopy (ARPES) study of the electron-doped cuprates Pr$_{1.3-x}$La$_{0.7}$Ce$_x$CuO$_4$ showing spin-glass, disordered AFM behaviors, and superconductivity at low temperatures and, by measurements with fine momentum cuts, found that the gap opens on the unfolded Fermi surface rather than the AFM Brillouin zone boundary. The gap did not show a node, following the full symmetry of the Brillouin zone, and its magnitude decreased from the zone-diagonal to ($\pi$,0) directions, opposite to the hole-doped case. These observations were reproduced by cluster dynamical-mean-field-theory (CDMFT) calculation, which took into account  electron correlation precisely within a (CuO$_2$)$_4$ cluster. The present experimental and theoretical results are consistent with the mechanism that electron or hole doping into a Mott insulator creates an in-gap band that are separated from the upper or lower Hubbard band by the pseudogap.
	\end{abstract}
	
	
	\maketitle

	\section{Introduction}
	One of the most unusual and intriguing features of the cuprate superconductors is the pseudogap phenomena. 
	So far, a variety of proposals have been made as the origin of the pseudogap. 
	Among them, phase-incoherent Cooper pairs \cite{Emery1995} and competing orders such as charge order \cite{Comin2016} and nematic order \cite{Hinkov2008,Lawler2010,Kivelson2019} have been extensively discussed. 
	Another scenario is based on the proximity to the Mott insulator, without relying on any competing orders. 
	Unbiased numerical calculations of the hole-doped Hubbard \af{and $t-J$ models} have shown that a pseudogap with the full symmetry of the Brillouin zone opens between the lower Hubbard band (LHB) and an in-gap band (IGB), which is split off the LHB in the entire momentum space \sakai{\cite{Civelli2005,Kyung2006,Tohyama2004,Sakai2010,Sakai2013}}. That is, upon hole doping, an electron at the LHB is {\it fractionalized} into two fermion components. 
	The two components are hybridizing with each other and generates a hybridization gap, which emerges as a pseudogap. The original LHB is thus split into the IGB and a fraction of the LHB~\cite{Yamaji2011,Imada2019,imada_review2021}. 
	
	In the above electron-fractionalization picture, the IGB mostly resides on the unoccupied side of the Fermi level ($E_\mathrm{F}$) except for the antinodal region, where a lower portion of the pseudogap is detectable by angle-resolved photoemission spectroscopy (ARPES). 
	Pseudogap structures above $E_\mathrm{F}$ to some extent have been studied by ARPES utilizing the temperature broadening of the Fermi-Dirac distribution function \cite{Yang2008,Yang2011}, but the major part of the pseudogap, which may extend up to several hundreds meV above $E_\mathrm{F}$, remains inaccessible in the hole-doped cuprates.  
	Meanwhile, electronic Raman scattering experiment, which covers both above and below $E_{\rm F}$, has been able to study the entire pseudogap energy region~\cite{Sakai2013}. Recently, resonant inelastic x-ray scattering (RIXS) has been used to access the empty states of hole-doped cuprates up to $\sim$1 eV above $E_\mathrm{F}$ and successfully detected electron fractionalization in the pseudogap state~\cite{Singh2022,Imada2021}. Electron-doped cuprates have a great advantage over the hole-doped ones in studying the pseudogap and the IGB since they are located below $E_\mathrm{F}$ and easily accessible by ARPES. 
	
	So far, the pseudogap of the electron-doped cuprates has been observed through the measurements of optical conductivity \cite{Onose2001,Wang2006}, scanning tunneling spectroscopy \cite{Zimmers2007b}, and ARPES \cite{Armitage2001b,Matsui2005a,Park2007,Matsui2007,Richard2007,Ikeda2009,Song2012,Horio2016}. 
	Its origin has been naturally attributed to antiferromagnetic (AFM) correlations since the spectral intensity is strongly suppressed around the ``hot spot", where the Fermi surface and the AFM Brillouin zone (BZ) boundary cross each other. 
	In fact, a neutron scattering study of Nd$_{2-x}$Ce$_x$CuO$_4$ (NCCO) by Motoyama \textit{et al.}~\cite{Motoyama2007} have shown that a static AFM long-range order exist in the range $x<0.14$ at low temperatures and that, above the N{\'e}el temperature $T_{\rm N}$, AFM correlation length $\xi_\mathrm{AFM}$ and the thermal de Broglie length $\xi_\mathrm{th}$ satisfy $\xi_\mathrm{AFM} \sim 2.6 \xi_\mathrm{th}$ at the pseudogap onset temperature $T^*$ in a wide doping range, supporting a connection between the AFM correlations and the pseudogap~\cite{Kyung2004}.
	
	ARPES studies of electron-doped cuprates have shown that a pseudogap opens at the ``hot spots'', where the Fermi surface and the AFM BZ boundary cross with each other~\cite{Armitage2001b,Matsui2005a,Matsui2007,JFHe2020}, consistent with AFM band folding. 
	However, some ARPES results do not agree with the band folding of the AFM ground state. 
	The gap in NCCO has been observed up to $x\sim 0.16$, beyond the AFM ordered phase, suggesting that short-range AFM order or dynamical AFM correlation can also open the pseudogap~\cite{JFHe2020}.
	Furthermore, ARPES studies by Park \textit{et al.}~\cite{Park2007,Park2013} have shown the absence of clear AFM band folding even at base temperature. 
	The pseudogap observed by ARPES has been reproduced by a theoretical model with disordered, short-range antiferromagnetism with a correlation length of only $\sim 4$ lattice spacings, even though the neutron-scattering study~\cite{Motoyama2007} suggests one order of magnitude larger correlation length. 
	As for the global momentum dependence of the pseudogap, the ARPES results reported by Matsui {\it et al.} \cite{Matsui2005a} and Ikeda {\it et al.} \cite{Ikeda2009} have indicated the reduction of pseudogap magnitude upon approaching ($\pi$, 0). This momentum dependence cannot be understood within the simple AFM band-folding picture, which predicts a constant gap magnitude on the AFM BZ boundary. Most of theoretical studies, either with weak-coupling \cite{Kyung2004} or strong-coupling treatment \cite{Senechal2004}, have not explicitly shown the momentum dependence of the AFM gap, whereas a variational Monte-Carlo study based on the two-dimensional $t$-$t'$-$t"$-$J$ model by Chou and Lee \cite{Chou2008} and a variational cluster-perturbation theory study of the Hubbard model by S\'en\'echal \textit{et al.}~\cite{SenechalPRL2005} have shown that the gap decreases as one goes from the hot spot to ($\pi$, 0). The failure of the simple AFM band-folding picture and the previous theoretical models to explain the salient features of the ARPES data should contain crucial information about the nature of the electron-electron and spin-spin correlations not only in the electron-doped cuprates but also in the hole-doped cuprates, and calls for further investigations.
	
	In this Article, we report on an ARPES study of the electron-doped cuprate Pr$_{1.3-x}$La$_{0.7}$Ce$_x$CuO$_4$ (PLCCO, $x = 0.02$) showing spin-glass and disordered AFM behaviors at low temperatures ($T_{\rm N} \simeq 140$ K, see Materials and Methods for the definition of $T_{\rm N}$). 
	The doped electron concentration  
	was estimated from Fermi surface area to be $n_{\rm FS}= 0.10\pm 0.01$.  This spin-glass state suggests close proximity to the phase boundary with superconducting states in the phase diagram~\cite{Kuroshima2003}. 
	The samples were protect-annealed under strongly reducing condition to efficiently remove excess oxygen \cite{Adachi2013}. We investigated the band structure along various directions in momentum space, and found that the pseudogap is indeed strongly momentum dependent and becomes invisible in the antinodal region. Even more importantly, the momentum position of the gap was found to deviate from the AFM BZ boundary around the antinode, which is incompatible with AFM pictures of the pseudogap. Qualitatively the same ARPES spectra were also obtained for a superconducting $x = 0.05$ sample ($n_{\rm FS}= 0.13\pm 0.01$) with $T_c = 24$ K showing a disordered AFM behavior ($T_{\rm N} \simeq 85$ K). These results are consistent with \af{cluster dynamical-mean-field theory (CDMFT)} calculation, which takes into account \af{short-range} electron correlation precisely and is relevant to the disordered AFM state, and support the picture that the dynamical correlation with the correlation length of a few lattice spacings is of direct relevance to the pseudogap formation for the electron-doped cuprates.
	
	\begin{figure}[t]
		\centering
		\includegraphics[width=0.48\textwidth]{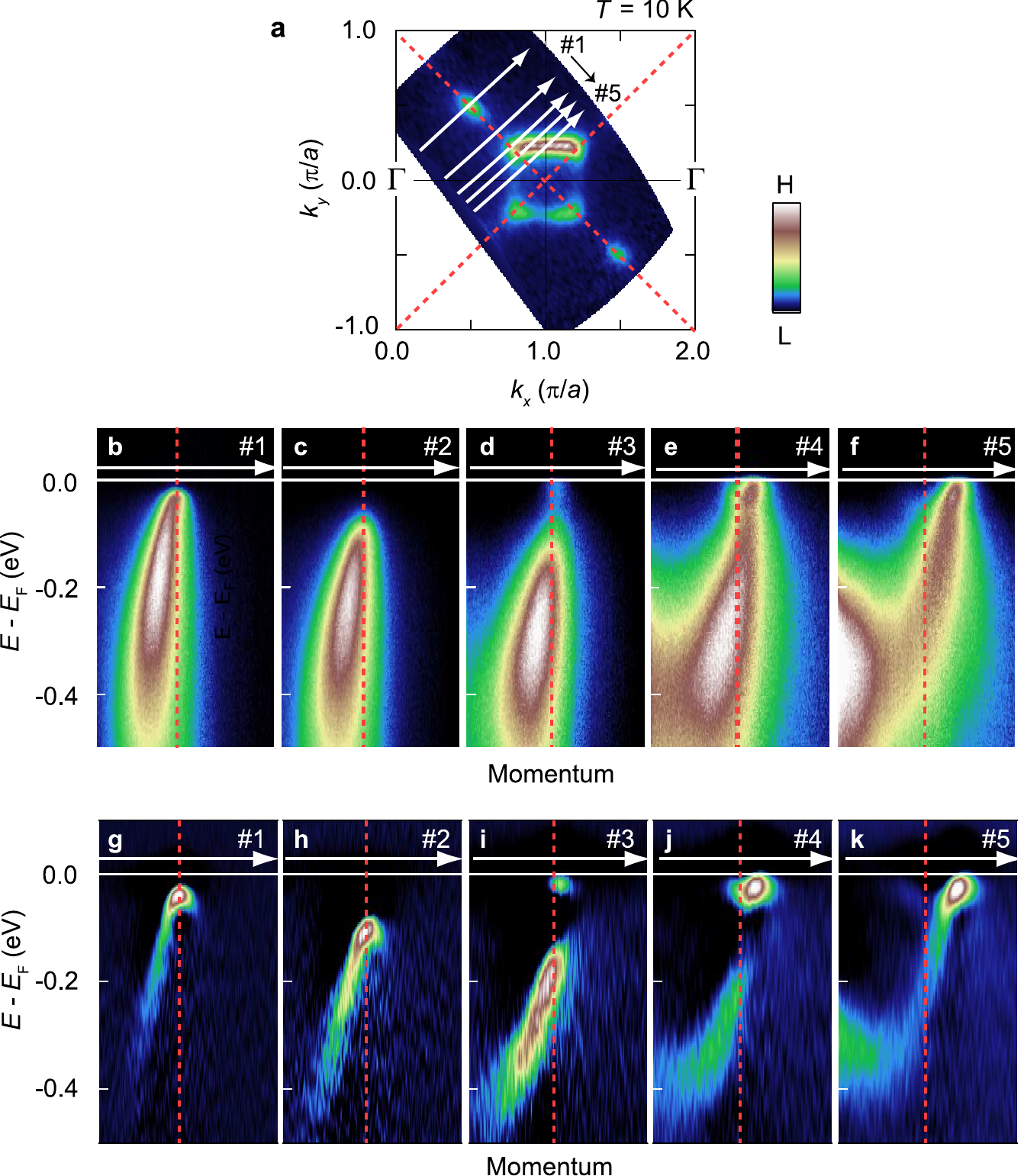}
		\caption{\textbf{ARPES spectra of a protect-annealed PLCCO (\boldmath{$x = 0.02$}) single crystal.} \textbf{a}, Fermi surface map measured at 10 K. Spectral intensity integrated within $\pm 20$ meV of $E_\mathrm{F}$ is plotted. \textbf{b-f}, Band images taken along the cuts indicated in (a). The antiferromagnetic (AFM) Brillouin-zone (BZ) boundary is shown by dashed red lines in each panel. \textbf{g-k}, Corresponding second derivatives with respect to energy. 
		}
		\label{AFPG_band}
	\end{figure}
	
	\section{Results and analysis}
	Figure~\ref{AFPG_band} shows the Fermi surface, band images, and their second derivatives with respect to energy of a protect-annealed PLCCO ($x = 0.02$) sample at 10~K. The band is gapped at $E_\mathrm{F}$ at the node (cut~\#1), and the gap below $E_\mathrm{F}$ increases in going from the node to the hot spot (cut~\#2). On approaching the antinode, the upper split band is lowered below $E_\mathrm{F}$ and produces a finite spectral intensity at and below $E_\mathrm{F}$. These features are characteristics of the electron-doped cuprates as have been reported in previous studies \cite{Matsui2005a,Park2007,Matsui2007,Richard2007,Ikeda2009,Song2012,Horio2016}.
	
	Now we look into more details of the band structure. At the node and the hot spot (cuts~\#1 and \#2), as the band disperses from higher binding energy toward $E_\mathrm{F}$, it disperses back to higher binding energy beyond the AFM BZ boundary, producing a local band maximum at the AFM BZ boundary (Figs.~\ref{AFPG_band}g and h) as if the band is folded by the ${\bf Q} = (\pi, \pi)$ AFM order. Once the upper band is lowered below $E_\mathrm{F}$ (cuts~\#3, \#4, and \#5), however, the lower band is no longer folded and disperses straightly across the AFM BZ boundary. Qualitatively consistent ARPES spectra were also obtained for a protect-annealed PLCCO sample with $x = 0.05$, which exhibits superconductivity as well as lower magnetic transition temperature (See Supplementary Note~1 and Supplementary Fig.~S1).

	\begin{figure}[t]
		\centering
		\includegraphics[width=0.48\textwidth]{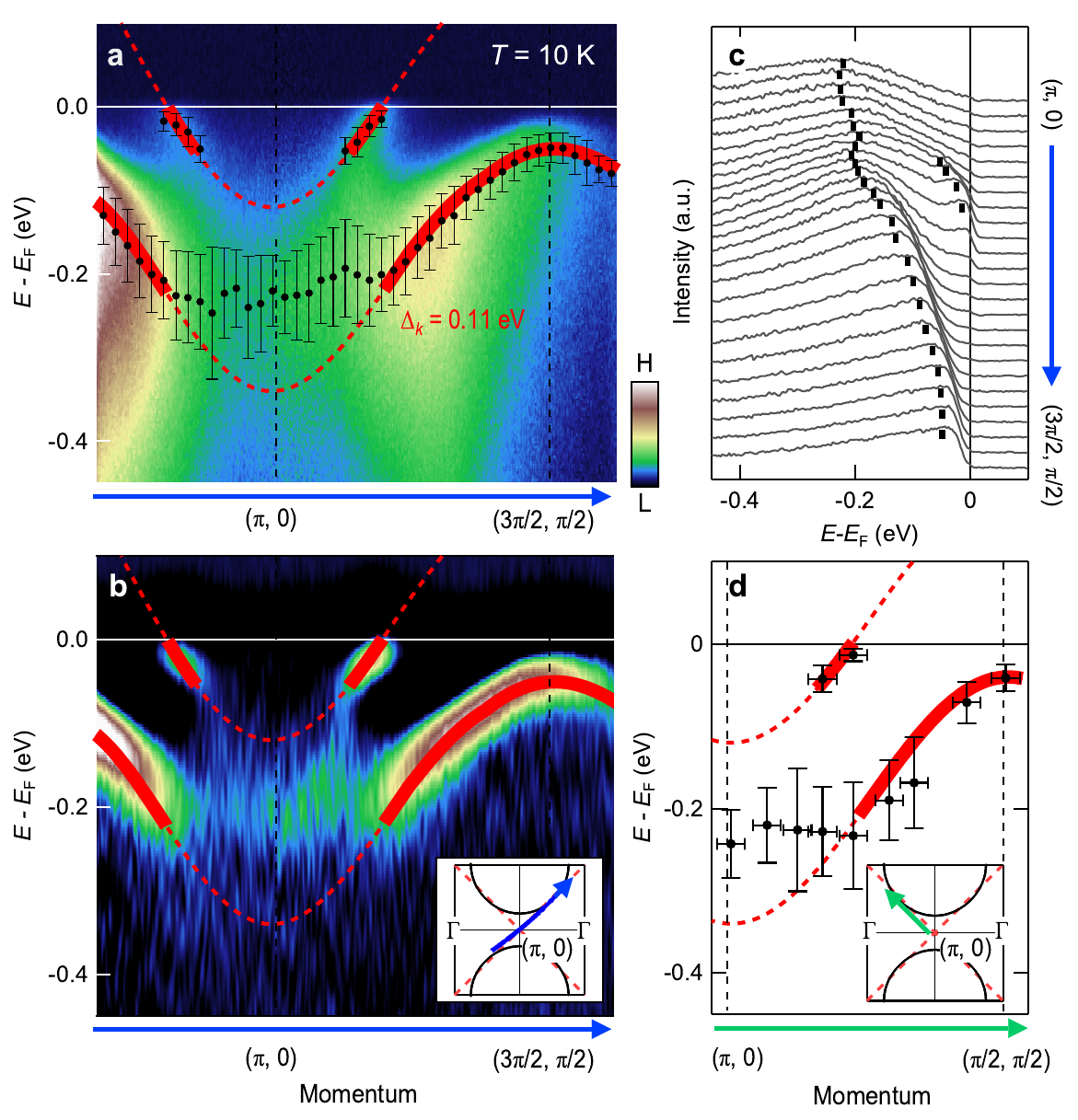}
		\caption{\textbf{Band splitting measured along the folded BZ boundary.} \textbf{a},\textbf{b}, Band image recorded at 10~K and its second derivative with respect to energy along the cut indicated in the inset of \textbf{b}. Peak positions of the EDCs are plotted as black dots in \textbf{a}. The error bar is defined by the lower-energy-edge width at half maximum ($\times 33$~\%). Band dispersions fitted to the mean-field-type model for a ${\bf Q} = (\pi, \pi)$ order with $\Delta_k = 0.11$ eV (red curves) are also plotted, in which the region relevant to the fitting is represented by solid curves (outside the antinodal region). \textbf{c}, EDCs plotted along the folded BZ boundary from ($\pi$, 0) to ($3\pi/2$, $\pi/2$) with black bars representing peak positions. \textbf{d}, Peak positions of EDCs from ($\pi/2$, $\pi/2$) to ($\pi$, 0) extracted from the measurements along several cuts parallel to the (0, 0)-($\pi$, $\pi$) direction. Band dispersions calculated using the simple AFM band-folding model with the same parameters as above are superimposed, indicating that the band dispersions shown in \textbf{b} and \textbf{d} are identical.}
		\label{AFPG_BZ}
	\end{figure}
	
	The experimentally obtained band image along the AFM BZ boundary, its second derivative with respect to energy, and energy distribution curves (EDCs) are shown in Figs.~\ref{AFPG_BZ}a--c. Starting from (3$\pi$/2, $\pi$/2), the peak position of the EDC disperses smoothly toward higher binding energy, but stops dispersing when the spectral intensity of the upper band emerges below $E_\mathrm{F}$. Once the upper band reaches the binding energy of $\sim 50$ meV, the magnitude of the gap decreases, the width of the quasi-particle (QP) peak increases, and the two peaks become broad and unresolved around ($\pi$, 0). 
	
	Low-energy spectra around ($\pi$, 0) are plotted in Figs.~\ref{AFPG_antinode}b--d, where one finds that the gap opens away from the AFM BZ boundary. This is in contrast to the nodal region where the band energy position is maximized near the AFM BZ boundary and thus the gap apparently opens around it (cuts~\#1 and \#2). The peculiar momentum dependence of the gap opening positions is summarized in Fig.~~\ref{AFPG_antinode}e [See Supplementary Fig.~S2 for the gap-position determination around ($\pi$, 0)].
	
	\subsection{AFM band-folding model}
\begin{figure}[t]
		\centering
		\includegraphics[width=0.47\textwidth]{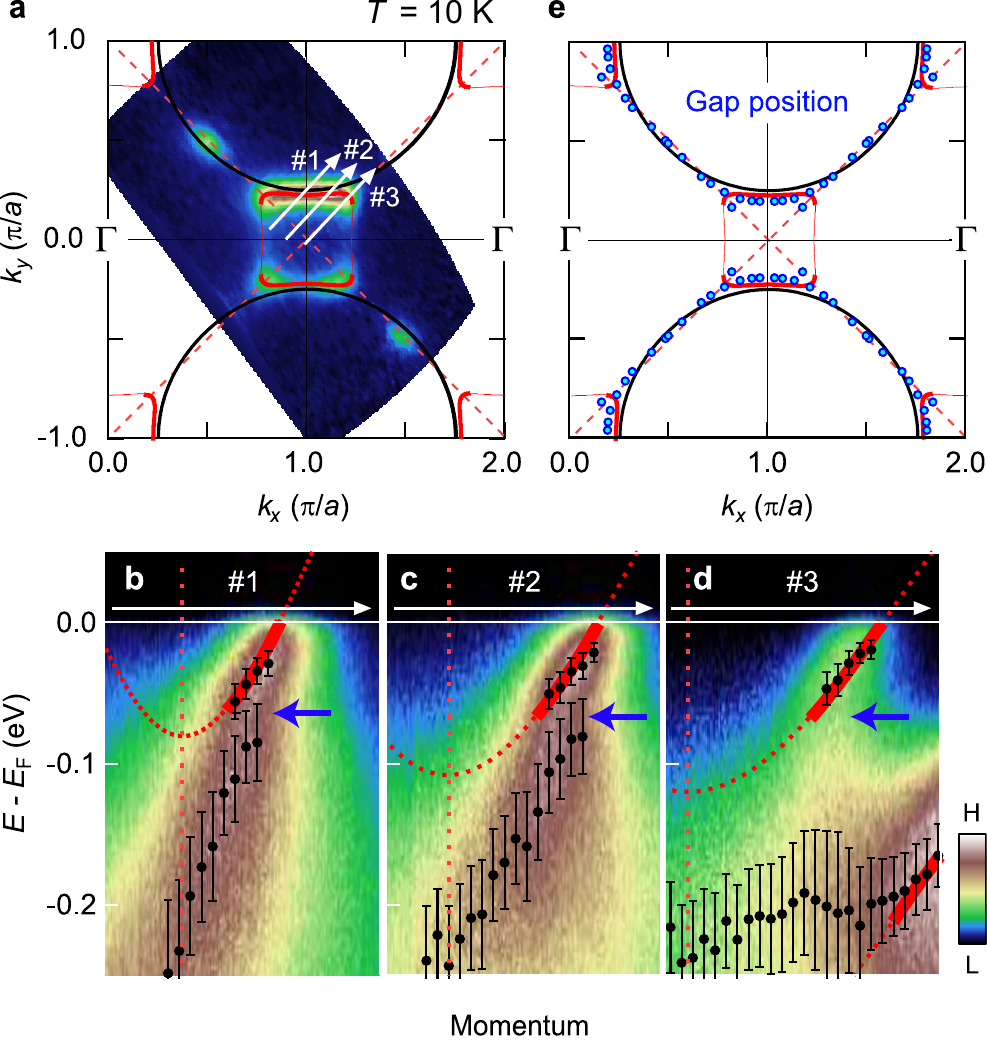}
		\caption{\textbf{Low-energy band structure around (\boldmath{$\pi$}, 0).} \textbf{a}, Fermi surface map at 10~K with the Fermi surface contour calculated using the AFM band-folding model with $\Delta_k = 0$ eV (black curves) and with $\Delta_k = 0.11$ eV (red curves). \textbf{b--d}, Band images along cuts~\#1--\#3 indicated in \textbf{a}, respectively. Peak positions of the EDCs are plotted with black dots. The definition of the error bar is the same as in Fig.~\ref{AFPG_BZ}. The bands calculated using the AFM band-folding model are also plotted. The lower bands obtained in the fitting shown in Figs.~\ref{AFPG_BZ}a and b disperse far out side of panels b--d. A gap opens away from the AFM BZ boundary as indicated by blue arrows. \textbf{e}, Momentum position of the gap opening. Near ($\pi$/2, $\pi$/2) where only the lower band is visible, it has been defined as the lower band top. In the antinodal part, it has been determined from the momentum-distribution-curve peak in between the upper and lower bands.}
		\label{AFPG_antinode}
	\end{figure}

	In order to describe these unusual band dispersions, we first try to employ mean-field-type AFM band-folding on the square lattice with a ${\bf Q} = (\pi, \pi)$ AFM order \cite{Matsui2005a,Park2007,Ikeda2009}, where transfer integrals between the nearest-neighbor ($t$), second-nearest-neighbor ($t'$), and third-nearest-neighbor ($t''$) Cu sites are considered. In the mean-filed picture, the potential difference between the two sub-lattices, 2$\Delta_k$, is independent of electron momentum \textbf{\textit{k}}, 
	which cannot explain the observed strong variation of the gap size along the AFM BZ boundary. Nevertheless, allowing for different $\Delta_k$ values for different \textbf{\textit{k}} regions within the AFM band-folding model, we could fit the experimental band dispersions with parameter values $t = 0.25$ eV, $t'/t = -0.15$, and $t''/t' = -0.5$ as shown in Figs.~\ref{AFPG_BZ}a, b, and \ref{AFPG_antinode}b-d by solid curves. Here, $\Delta_k=0.11$ eV yielded the best fit around (3$\pi$/2, $\pi$/2) and $\Delta_k \sim 0$ eV around ($\pi$, 0). The doped electron concentration $n_\mathrm{FS}$ was estimated from the Fermi surface area to be 0.104, which is considerably larger than the nominal Ce concentration of $x=0.02$. 
	Such excess electrons compared to the nominal Ce content are due to oxygen vacancies resulting from strong reduction as revealed by recent ARPES studies \cite{Horio2016,Song2017,HorioPRB2018}. Note that the rapid change of $\Delta_k$ cannot be justified within the mean-field-type band-folding picture, and calls for need to go beyond the band-folding picture.

	In order to see possible effects of photoionization matrix elements, we have also examined the band dispersion along the folded BZ boundary in a different direction, from ($\pi$, 0) to ($\pi/2$, $\pi/2$). As plotted in Fig.~\ref{AFPG_BZ}d, the observed band dispersions are identical to the dispersions from ($\pi$, 0) to (3$\pi$/2, $\pi$/2) (Figs.~\ref{AFPG_BZ}a and b). The present results thus highlight the momentum-dependent band splitting in the electron-doped cuprates: The large pseudogap and sharp spectra around ($\pi/2$, $\pi/2$) and the reduced pseudogap and broad spectra near ($\pi$, 0). This momentum dependence of the pseudogap magnitude is opposite to the hole-doped cuprates, where the pseudogap is large around ($\pi$, 0) and is apparently closed around ($\pi/2$, $\pi/2$) below $E_\mathrm{F}$.
	
	Here, we also note the difficulty of the AFM band-folding model in explaining the AFM state of the undoped and lightly electron-doped cuprates as follows: If the observed pseudogap arose from AFM band folding, one would be able to make a rough estimate of the AFM gap \mhh{2}$\Delta_k$ for the undoped compound by extrapolating the doping dependence. Although the lowest doping level we could achieve with the protect-annealed samples was $n_{\rm FS}\sim$ 0.09 as measured by the Fermi surface area, extrapolating the data plotted in Fig. S6 (including the results from as-grown samples) to $n_{\rm FS}\sim$ 0 yields $2\Delta_k \sim 0.4$~eV. 
	This gap value is much smaller than the $\sim 1$~eV band gap of the parent insulator~\cite{Armitage2002}.

	\subsection{Beyond the AFM band-folding model}	
	Although the above AFM band-folding analysis based on the tight-binding model helps us to characterize the momentum-dependent pseudogap, it requires the abrupt collapse of the $\Delta_k$ parameter around ($\pi$, 0), indicating the breakdown of the mean-field-type band-folding picture. Furthermore, the key aspects of the experimental results are the absence of band folding and of the spectral-weight suppression across the folded BZ boundary in the antinodal lower band (Figs.~\ref{AFPG_band}f and g). Even more important is the gap opening away from the folded BZ boundary in the antinodal region (Figs. \ref{AFPG_antinode}b--d). While the gap positions nearly trace the AFM BZ boundary around ($\pi/2$, $\pi/2$), they deviate from the AFM BZ boundary as the cut approached ($\pi$, 0). The gap positions in {\bf k} space plotted in Fig.~\ref{AFPG_antinode}e follows the original Fermi surface before the AFM band folding, on which the gossamer Fermi surface~\cite{KJXu2023} and the electron pocket are formed and are smoothly connected, rather than the AFM BZ boundary.  These spectral features are incompatible with the slow, long-range AFM correlation scenarios and suggest the importance of more short-ranged, dynamical AFM correlation in creating the pseudogap.

	\subsubsection{CDMFT and electron fractionalization}
\begin{figure}[t]
		\centering
		\includegraphics[width=0.47\textwidth]{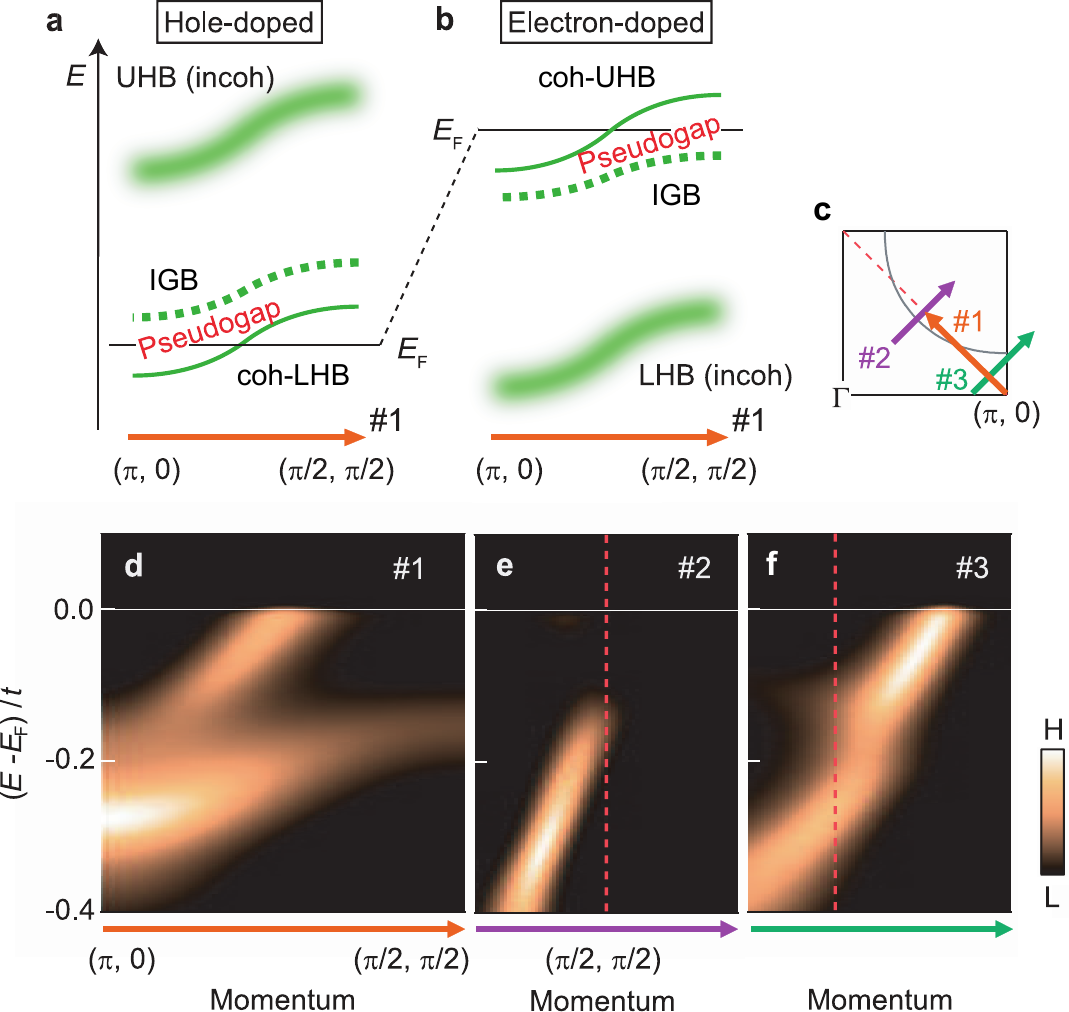}
		\caption{
			\textbf{Formation of the pseudogap in the hole- and electron-doped cuprates and cluster dynamical-mean-field-theory (CDMFT) calculation for the electron-doped cuprates .} \textbf{a,b}, Schematic band dispersions along the AFM BZ boundary [cuts~\#1 and \#2 in panel \textbf{c}] for the hole-doped and electron-doped cuprates with the pseudogap without a node, respectively. IGB, coh-LHB, and coh-UHB denote the in-gap band, the coherent part of the lower Hubbard band and that of the upper Hubbard band, respectively, in the electron-fractionalization picture. Note that the LHB in the single-band Hubbard model is equivalent to the Zhang-Rice singlet band in the three-band Hubbard model after appropriate parameter conversion~\cite{Sheshadri2023}. In the hole-doped (electron-doped) cuprates, the LHB (UHB) is fractionalized into the coh-LHB (coh-UHB) and the IGB. CDMFT calculations yield smaller pseudogap around ($\pi/2$, $\pi/2$) [around ($\pi$, 0)] for hole (electron)-doped cuprates. \textbf{c}, Momentum cuts for \textbf{d}, \textbf{e}, and \textbf{f}. \textbf{d-f}, Spectral function calculated by CDMFT along cuts~\#1-\#3 indicated in panel \textbf{c}. See text for details of the calculation.}
		\label{DMFT}
	\end{figure}

	One scenario to explain the overall band structure without assuming the long-range AFM order is the nodeless pseudogap proposed by Sakai {\it et al.}~\cite{Sakai2010,Sakai2013}, based on a CDMFT calculation \cite{Kotliar2001} on the Hubbard model. They
	have shown that the pseudogap of the hole-doped cuprates does not show a node and opens in the entire momentum space \af{while, around $(\pi/2,\pi/2)$, the gap resides on the unoccupied side and thus the Fermi surface survives.} As schematically shown in Fig.~\ref{DMFT}a, hole doping into the Mott insulator splits the LHB into the coherent part of the LHB (coh-LHB) and the IGB. The IGB and coh-LHB reside above $E_\mathrm{F}$ in most of momentum space while the coh-LHB disperses to below $E_\mathrm{F}$ around ($\pi$, 0). Since unoccupied states cannot be detected by ARPES, the pseudogap of hole-doped cuprates has been observed only in the antinodal region. 
	
	In the case of electron-doped cuprates, as shown in Fig.~\ref{DMFT}b, the chemical potential is shifted to the bottom of the upper Hubbard band (UHB), and the UHB is fractionalized into the coherent part of the UHB (coh-UHB) and the IGB just below the UHB. Both of the split UHBs are occupied in the region around ($\pi$, 0) in contrast to the hole-doped case while only the IGB is occupied in the nodal region, reproducing the experimental results. The fully gapped structure in the nodal region cannot be accounted for by the $d$-wave pseudogap with particle-hole symmetry expected from the $d$-symmetry pairing precursor origin of the pseudogap. Nevertheless, it should be noted that superconducting fluctuations can open a normal-state gap of $d$-symmetry on the lower energy scale of  $\sim$10 meV in underdoped NCCO ($x<$0.1)~\cite{KJXu2024}, on top of the pseudogap of order $\sim$100 meV reported here.
	
	\subsubsection{CDMFT calculation compared with ARPES}
	Following the above consideration, we have carried out 2 $\times$ 2 CDMFT calculations of the Hubbard model for the paramagnetic state using parameters $n=0.11$, $t'=-0.3t$, and $U=8t$, where $n$ ($>0$) is the doped electron concentration and $U$ is the on-site Coulomb repulsion. $n$ has been set close to the value deduced from the observed Fermi surface area $n_\mathrm{FS} = 0.104$. The $t'/t$ value has been derived from density-functional-theory calculation \cite{Ikeda2009}. 
	The $U/t$ value of 8 is in accordance with the theoretical estimate for hole-doped cuprates ($U/t=$ 7-10) using the multi-scale {\it ab initio} scheme for correlated electrons (MACE)~\cite{Moree2022}, with considering a similar or slightly smaller $U/t$ value for electron-doped cuprates \cite{Jang2016}. Note that the choice of a smaller $U/t = 6.5$, which is close to the values adopted in previous theoretical studies~\cite{Kyung2004,YuanPRB2005,Weber2010a} also yields qualitatively the same result with a somewhat smaller pseudogap (See Supplementary Fig.~S3). The obtained spectral function has been multiplied by the Fermi-Dirac function and resolution-broadened for comparison with the ARPES spectra.

	The spectral function calculated along the AFM BZ boundary plotted in Fig.~\ref{DMFT}d indicates that the pseudogap shrinks on approaching ($\pi$, 0), which explains the experimental results (Fig.~\ref{AFPG_BZ}). The electron-concentration dependence of the calculated spectra is also qualitatively consistent with the experimental results (See Supplementary Figs.~S4--S6). Most remarkably, the calculated antinodal spectrum in Fig.~\ref{DMFT}f reproduces the essential characteristics (cut~\#5 in Fig.~\ref{AFPG_band}, cuts~\#1 and \#2 in Fig.~\ref{AFPG_antinode}) that the lower band is not folded at the AFM BZ boundary but disperses across it, and the gap opens apart from the AFM BZ boundary. The overall momentum positions of the gap opening (Fig.~\ref{AFPG_antinode}e) is also reproduced by the calculation satisfactorily (see Supplementary Fig.~S7). The antinodal spectra shown in Figs.~\ref{AFPG_antinode}b--d indicate that the upper band loses intensity and the pseudogap opens once it reaches the binding energy of $\sim 50$ meV. These behaviors of the fractionalized bands cannot be explained by the AFM band-folding picture. In the fractionalized fermion picture, this can be interpreted as a consequence of hybridization between the UHB and a ``dark/hidden fermion" \cite{Sakai2016}. The ``dark fermion" is a phenomenologically introduced virtual fermion that describes the electron fractionalization. In \af{theory}, the dark-fermion dispersion is the trace of the maximal point of the imaginary part of the self-energy or the zero of the Green function \cite{Sakai2016}. The calculated nodal spectra (Fig.~\ref{DMFT}e) lose intensity across the AFM BZ boundary, in agreement with the corresponding ARPES spectra (cut~\#1 in Fig.~\ref{AFPG_band}).

\begin{figure}[t]
			\centering
			\includegraphics[width=0.49\textwidth]{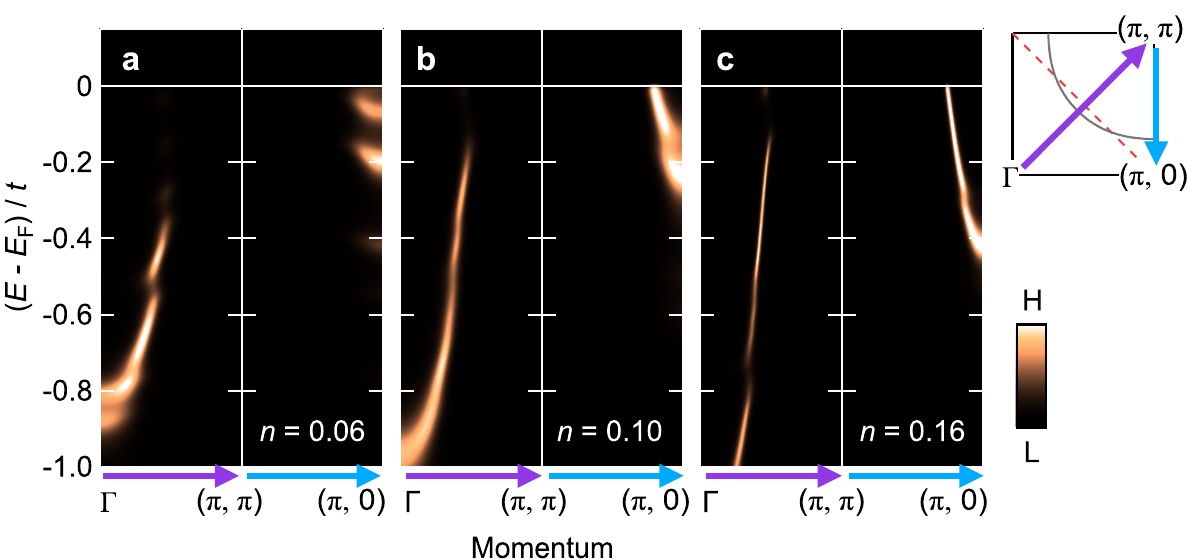}
			\caption{
				\textbf{Evolution of the pseudogap by electron doping.} Spectral function calculated by CDMFT along the cuts indicated in the rightmost panel for the doped electron concentration $n$ of \textbf{a,} 0.06, \textbf{b,} 0.10, and \textbf{c,} 0.16, respectively. The spectra have been multiplied by the Fermi-Dirac function.
			}
			\label{doping}
		\end{figure}

	The consistency between the paramagnetic CDMFT result and the ARPES spectra is i) the nodeless pseudogap, ii) the electron-hole asymmetry, iii) the reduction of the gap in the antinodal region, and iv) the gap opening not on the AFM BZ boundary but away from it. Points i) and ii) can be explained by the AFM band-folding picture and point iii) by the variational Monte-Carlo~\cite{Chou2008} 
	and variational cluster-perturbation theory studies~\cite{SenechalPRL2005}, 
	respectively, for the AFM and coexisting AFM and $d$-wave superconducting states. However, our crucial point iv) 
	is not reproduced by calculations assuming AFM order. 
	Even 
		spin-wave theory beyond the mean-field AFM ordered state
		~\cite{Marsiglio1991} cannot account for the present ARPES results. 
		Note that in the present CDMFT calculations, AFM order is absent while all the short-range charge and spin correlations within the $2 \times 2$ cluster are fully taken into account. 
		Our result reproducing all the salient features of the ARPES spectra by the $2 \times 2$ {\it paramagnetic} CDMFT calculations
		suggests that the Mott physics is at the origin of the pseudogap of electron-doped cuprates through the enhanced electron correlations without long-range AFM correlations, as in the case of the hole-doped cuprates~\cite{MaierPRB2002,Civelli2005,Kyung2006,FerreroPRB2009,Sakai2010,Sakai2013}.
		The present results thus indicate the formation of the particle-hole asymmetric pseudogap in the cuprates of both carrier types, without relying on competing instabilities towards charge~\cite{Comin2016}, AFM~\cite{Motoyama2007}, or nematic order~\cite{Hinkov2008,Lawler2010}. The momentum dependence of the pseudogap (Fig.~\ref{AFPG_BZ}) with an indirect gap (cut~\#5 in Fig.~\ref{AFPG_band}, cuts~\#1 and \#2 in Fig.~\ref{AFPG_antinode}) shows a remarkable resemblance with the Mott gap, although the magnitude is much smaller, as one sees in Fig.~\ref{DMFT}a (See Supplementary Fig.~S4 for the calculated spectral function including the unoccupied side), in contrast to the direct gap predicted for the AFM gap. The CDMFT calculations also \mh{qualitatively} reproduce the enhanced pseudogap at low electron doping (Fig.~\ref{doping}) observed in the previous ARPES study~\cite{Armitage2002}\mh{, while on the quantitative level the persistence of the pseudogap on the overdoped side seems to be overestimated within the current model and parameter choice}.

		\section{Discussion}
		Historically, pseudogap mechanisms have been proposed for hole- and electron-doped cuprates in different ways.  The AFM correlation length $\xi_\mathrm{AFM}$ is typically larger for the electron-doped cuprates~\cite{YamadaPRB1998,Motoyama2007}. Also, the connection between $\xi_\mathrm{AFM}$ and the thermal de Broglie length $\xi_\mathrm{th}$ observed at $T = T^*$ for the electron-doped cuprate~\cite{Motoyama2007,Kyung2004} does not hold on the hole-doped side. These contrasting magnetic properties may point toward different origins of the pseudogap between the hole- and electron-doped cuprates. On the other hand, the present study has focused on a different aspect in a more direct way, that is, the momentum dependence of the pseudogap opening. The absence of band folding and the deviation of the gap position from the AFM BZ boundary in the antinodal region rather suggest a mechanism that does not rely on AFM correlations with long $\xi_\mathrm{AFM}$ 
		for the non-superconducting $x=0.02$ ($n_\mathrm{FS}=0.10$) sample as well as the superconducting $x=0.05$ ($n_\mathrm{FS}=0.12$) sample with a lower magnetic transition temperature (See Supplementary Note 1 and Supplementary Fig.~S1). This conclusion may appear at odds with substantial $\xi_\mathrm{AFM}$ of 20--50 lattice spacings reported from the neutron scattering studies at the boundary between the AFM and superconducting phases~\cite{WilsonPRL2006,WilsonPRB2006} \mh{where similar magnetic properties to those of the present samples are expected~\cite{Fujita2003}}. However, the influence of AFM correlation on the single-particle spectral function is not trivial in the cuprates, as exemplified by stripe-ordered hole-doped cuprates. Even though \LSCO\ and \LBCO\ at the 1/8 doping possess $\xi_\mathrm{AFM} \sim 25$~\cite{YamadaPRB1998} and $\sim 150$ lattice spacings~\cite{HuckerPRB2011} at low temperatures, respectively, band structure observed by ARPES remains unaffected by the stripe order~\cite{HeNatPhys2009}. Based on the remarkable agreement between the present ARPES spectra and the CDMFT calculation, we suggest that strong correlations in the proximity of the Mott insulator supplemented by the highly dynamical (or disordered) AFM correlations with $\xi_\mathrm{AFM}$ of a few lattice spacings involving singlet correlation rather than low-energy AFM correlations with large $\xi_\mathrm{AFM}$ as the primary origin of the pseudogap, as has been proposed for the hole-doped cuprates~\cite{MaierPRB2002,Civelli2005,Kyung2006,FerreroPRB2009,Sakai2010,Sakai2013}. Still, at present, it is elusive how $\xi_\mathrm{AFM}$ is connected to $T^*$ when the pseudogap is driven by strong electron correlations. 
		
		Whereas the aforementioned variational Monte-Carlo study~\cite{Chou2008} 
		and the variational cluster-perturbation theory study~\cite{SenechalPRL2005} 
		were based on AFM order, our CDMFT calculation has been conducted for the paramagnetic state with short-range AFM and singlet correlations. Despite differences between these and our calculations, qualitative similarity in the momentum dependences except for the existence/absence of band folding suggests a common mechanism of the reduced band splitting in the antinodal region. Since both of the split bands are located above $E_\mathrm{F}$ in the nodal region of the hole-doped case (coh-LHB and IGB) and below $E_\mathrm{F}$ in the antinodal region of the electron-doped case (coh-UHB and IGB), these momentum regions do not gain the energy of the electron system by the pseudogap opening. This accounts for the diminished pseudogap in the different momentum regions between both cases. The CDMFT calculations thus explain the key experimental differences between the pseudogaps of the hole- and electron-doped cuprates; the enhanced pseudogap around $(\pi,0)$ in the hole-doped cuprates and the diminishing pseudogap in the same region in the electron-doped cuprates. This contrast between the hole- and electron-doped sides simply arises from the common dispersions of the upper and lower Hubbard bands, where the nodal part has higher energy than the antinodal part. The enhanced gap in the nodal region of the electron-doped cuprates is also \af{interpreted as} a stronger hybridization strength with the ``dark fermion"~\cite{Sakai2016}. It should be noted that this momentum dependence is clearly different from that of the precursor of $d$-wave superconductivity, which should diminish as one approaches the nodal region. 

		
		\section{Conclusion}
		The agreement between the CDMFT calculation and the ARPES experiment suggests that the 
		fractionalization of the UHB induced by electron doping into the Mott insulator is at the origin of the pseudogap formation in the electron-doped cuprates. The present work supports the scenario that the pseudogap of cuprates is driven by strong electron correlation itself and hence has an intimate link to the Mott insulating state rather than competing orders.
	
	\vspace{5mm}
	
	\vspace{5mm}
	\noindent
	\textbf{Methods}\\
	\textbf{Sample preparation:}
	Single crystals of PLCCO with $x$ = 0.02 were synthesized by the traveling-solvent floating-zone method. First the sample was protect-annealed at 850 $^\circ$C for 24 hours and then annealed at 400 $^\circ$C for 48 hours. Annealing generally suppresses the QP scattering rate~\cite{Xu1996} and sharpens ARPES spectra~\cite{Richard2007,Song2012,Horio2016}, which enables us to determine the band structure precisely. The annealed $x=0.02$ sample showed metallic behavior at high temperatures, but was not superconducting. According to muon-spin-rotation ($\mu$SR) measurements, the sample showed fast depolarization and hence disordered antiferromagnetism below $T_\mathrm{N}=$ 140 K. The ordering temperature was defined as the temperature where the magnetic volume fraction, which is determined by the magnitude of the exponential-type muon depolarization, exceeds 50 \%. This temperature, $T_\mathrm{N1}$, has been demonstrated to coincide with the temperature below which magnetic Bragg peaks emerge in neutron scattering measurements~\cite{Fujita2003}. Furthermore, the magnetic susceptibility showed difference between field cooling and zero-field cooling below $\sim 15$~K, indicative of a spin-glass state \cite{Kuroshima2003} (See Supplementary Note 2 and Supplementary Fig.~S8). The spin-glass behavior suggests that the sample is close to the boundary between the AFM and superconducting phases in the phase diagram. Note that the $\mu$SR measurements have been conducted on samples annealed at slightly lower temperatures. Specifically, they were protect-annealed at 825~$^\circ$C for 24 hours and then annealed at 400~$^\circ$C for 48 hours. Judging from the shorter $c$-axis lattice parameter, the removal of impurity apical oxygen atoms \cite{Radaelli1994} was more thoroughly made for the samples for the ARPES and magnetic susceptibility measurements, and hence the AFM order was more strongly suppressed. \\
	
	\noindent
	\textbf{ARPES measurements:} 
	ARPES measurements were carried out at beamline 5-4 of Stanford Synchrotron Radiation Lightsource. Circularly polarized light with $h\nu = 16.5$~eV was used. The total energy resolution was set at 15~meV. The sample was cleaved {\it in-situ} under pressure better than $3 \times 10^{-11}$~Torr and measured at 10~K. \\
	
	\noindent
	\textbf{CDMFT calculation:} 
	CDMFT calculation was performed with $2 \times 2$ interacting cluster coupled to eight bath sites. The cluster impurity problem was solved with the exact diagonalization method at zero temperature. The spectra were calculated with an energy-broadening factor 0.05$t$ and a cumulant-interpolation technique \cite{Stanescu2006} in momentum space. 
	
	\vspace{5mm}
	\noindent
	\textbf{Acknowledgements} \\
	Fruitful discussion with T.~K.~Lee, D.~Song, and C.~Kim is gratefully acknowledged. We thank M.~Ikeda for collaboration in the early stage of this work and J.~Xu and K.~Koshiishi for help in the ARPES measurements. The ARPES measurements were performed at the Stanford Synchrotron Radiation Lightsource, operated by the Office of Basic Energy Science, US Department of Energy. This work was supported by Grants-in-aid from JSPS (grant Nos.
	~JP16K05458, JP17H02915, JP17K14350, JP19K03741, JP22K13994, and JP22K03535) and from MEXT (grant Nos. JP22H05111, JP22H05114). 
	This work was also supported by MEXT as “Program for Promoting Researches on the Supercomputer Fugaku” (Basic Science for Emergence and Functionality in Quantum Matter - Innovative Strongly Correlated Electron Science by Integration of Fugaku and Frontier Experiments - JPMXP1020200104) and JPMXP1020230411 together with computational resources of supercomputer Fugaku provided by the
	RIKEN Center for Computational Science (Project ID: hp200132,
	hp210163, and hp220166). 
	A.F. acknowledges the support from the Yushan Fellow Program under the Ministry of Education of Taiwan.
	
	\vspace{5mm}
	\noindent
	\textbf{Author contributions} \\
	M. Horio, H.S., and Y.N. performed ARPES measurements with the assistance of M. Hashimoto. ARPES endstation was maintained by M. Hashimoto., D.L., and Z.-X.S. M. Horio analyzed the data. S.S. and M.I. performed CDMFT calculations. T.O., T.K., T.A., and Y.K. synthesized and characterized single crystals. M. Horio and A.F. wrote the manuscript with suggestions by S.S., M.I., T.A., and Y.K. All authors contributed to the scientific planning and discussions. A.F. was responsible for overall project direction and planning.
	
	\vspace{5mm}
	\noindent
	\textbf{Competing interests} \\
	The authors declare no competing interest.

	%

\end{document}


\title{Supplementary Information \\  Pseudogap in electron-doped cuprates: Strong correlation leading to band splitting}

	\author{M.~Horio}
\email{mhorio@issp.u-tokyo.ac.jp}
\affiliation{Department of Physics, University of Tokyo, Bunkyo-ku, Tokyo 113-0033, Japan}
\affiliation{Institute for Solid State Physics, The University of Tokyo, Kashiwa, Chiba 277-8581, Japan}

\author{S.~Sakai}
\email{shiro.sakai@riken.jp}
\affiliation{Center for Emergent Matter Science, RIKEN, Wako, Saitama 351-0198, Japan}

\author{H.~Suzuki}
\affiliation{Department of Physics, University of Tokyo, Bunkyo-ku, Tokyo 113-0033, Japan}
\affiliation{Frontier Research Institute for Interdisciplinary Sciences, Tohoku University, Sendai 980-8578, Japan}

\author{Y.~Nonaka}
\affiliation{Department of Physics, University of Tokyo, Bunkyo-ku, Tokyo 113-0033, Japan}

\author{M.~Hashimoto}
\affiliation{Stanford Synchrotron Radiation Lightsource, SLAC National Accelerator Laboratory, Menlo Park, California 94305, USA}

\author{D.~Lu}
\affiliation{Stanford Synchrotron Radiation Lightsource, SLAC National Accelerator Laboratory, Menlo Park, California 94305, USA}

\author{Z-.X.~Shen}
\affiliation{Stanford Synchrotron Radiation Lightsource, SLAC National Accelerator Laboratory, Menlo Park, California 94305, USA}

\author{T.~Ohgi}
\affiliation{Department of Applied Physics, Tohoku University, Sendai 980-8579, Japan}

\author{T.~Konno}
\affiliation{Department of Applied Physics, Tohoku University, Sendai 980-8579, Japan}

\author{T.~Adachi}
\affiliation{Department of Engineering and Applied Sciences, Sophia University, Tokyo 102-8554, Japan}

\author{Y.~Koike}
\affiliation{Department of Applied Physics, Tohoku University, Sendai 980-8579, Japan}

\author{M.~Imada}
\email{imada@g.ecc.u-tokyo.ac.jp}
\affiliation{Department of Engineering and Applied Sciences, Sophia University, Tokyo 102-8554, Japan}
\affiliation{Research Institute for Science and Engineering, Waseda University, Shinju-ku, Tokyo 169-8555, Japan}
\affiliation{Toyota Physical and Chemical Research Institute, Nagakute, Aichi 480-1192, Japan}

\author{A.~Fujimori}
\email{fujimori@phys.s.u-tokyo.ac.jp}
\affiliation{Department of Physics, University of Tokyo, Bunkyo-ku, Tokyo 113-0033, Japan}
\affiliation{Center for Quantum Science and Technology and Department of Physics, National Tsing Hua University, Hsinchu 30013, Taiwan}

\maketitle

\onecolumngrid

\section*{Supplementary Notes}

\section*{Supplementary Note 1: Physical properties and ARPES spectra of Pr$_{1.3-x}$La$_{0.7}$Ce$_x$CuO$_4$ (\boldmath{$x = 0.05$}) sample}
Single crystals of PLCCO with $x$ = 0.05 were synthesized by the traveling-solvent floating-zone method. First the sample was protect-annealed at 800 $^\circ$C for 24 hours and then annealed at 400 $^\circ$C for 48 hours. Thus annealed crystal showed superconductivity at $T_\mathrm{c} = 24$ K. The magnetic ordering temperature determined from $\mu$SR spectra was 85 K, which is lower than that of the $x=0.02$ annealed sample. ARPES spectra of the $x=0.05$ sample are displayed in Supplementary Fig.~S\ref{PLCCO005}. They are qualitatively consistent with those obtained from the $x=0.02$ sample.

\section*{Supplementary Note 2: Physical properties of protect-annealed Pr$_{1.3-x}$La$_{0.7}$Ce$_x$CuO$_4$ (\boldmath{$x = 0.02$}) samples}
\begin{figure}[h]
	\begin{center}
		\includegraphics[width=0.75\textwidth]{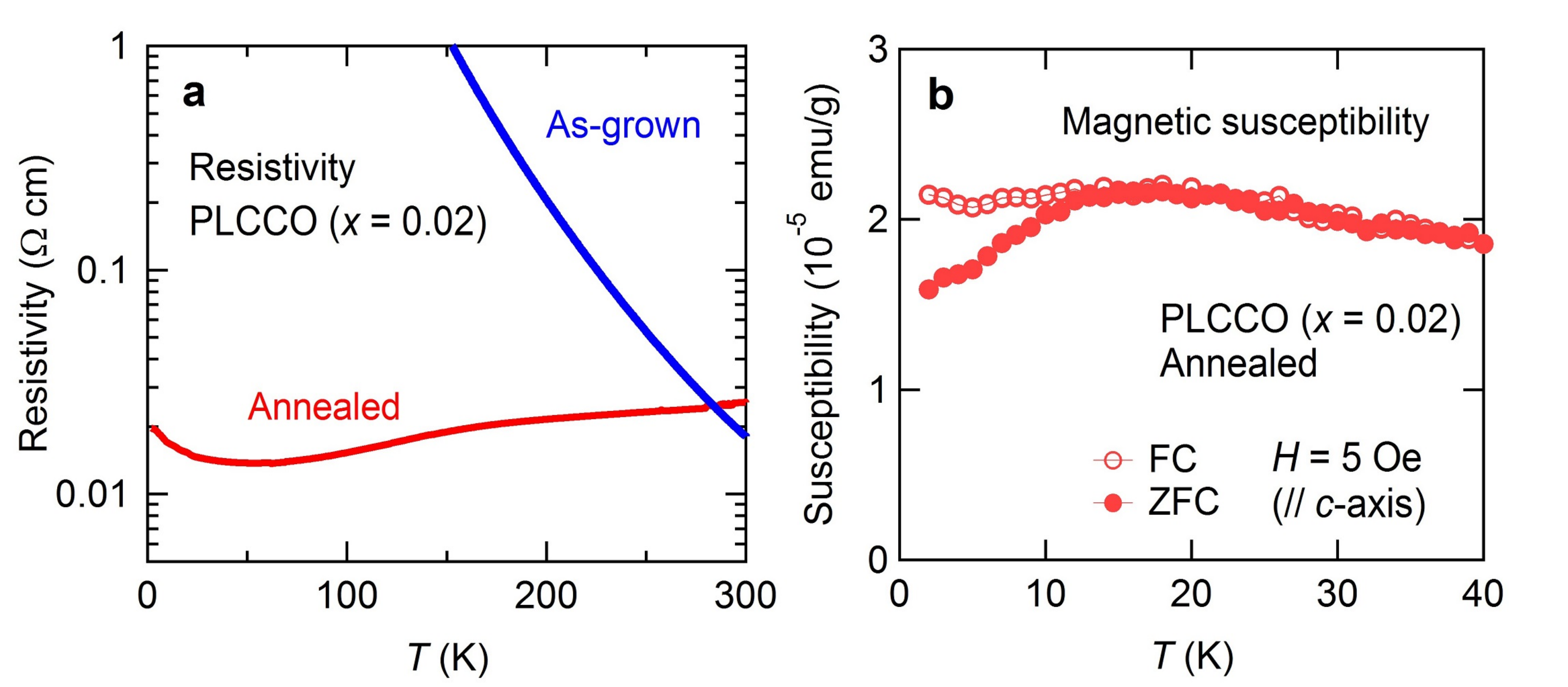}
	\end{center}
	\caption{\textbf{Physical properties of PLCCO (\boldmath{$x=0.02$}) single crystals.} \textbf{a}, Resistivity of as-grown and protect-annealed PLCCO ($x=0.02$) samples. \textbf{b}, Magnetic susceptibility of the annealed PLCCO ($x=0.02$) sample measured with field cooling (FC) and zero-field cooling (ZFC).}
	\label{sample}
\end{figure}

Supplementary Fig.~S\ref{sample}a shows the resistivity of as-grown and annealed Pr$_{1.3-x}$La$_{0.7}$Ce$_x$CuO$_4$ (PLCCO, $x = 0.02$) single crystals. While the as-grown sample is insulating, the protect-annealed sample shows metallic behavior down to $\sim 50$ K, probably due to the removal of impurity apical oxygen atoms. According to muon-spin-rotation ($\mu$SR) measurements, the rotation of muons were observed for both samples at low temperatures. The magnetic ordering temperature determined from $\mu$SR spectra, defined as a temperature where magnetic volume fraction develops to 50 \%, is 240 K and 140 K for the as-grown and annealed samples, respectively. The $\mu$SR spectra suggest that the AFM order in annealed PLCCO ($x = 0.02$) is not long ranged. The magnetic susceptibility of the annealed sample exhibits hysteresis as shown in Supplementary Fig.~S\ref{sample}b, which is indicative of a spin-glass behavior \cite{Kuroshima2003}.



\newpage

\section*{Supplementary Figures}

\begin{figure}[h]
\begin{center}
\includegraphics[width=0.99\textwidth]{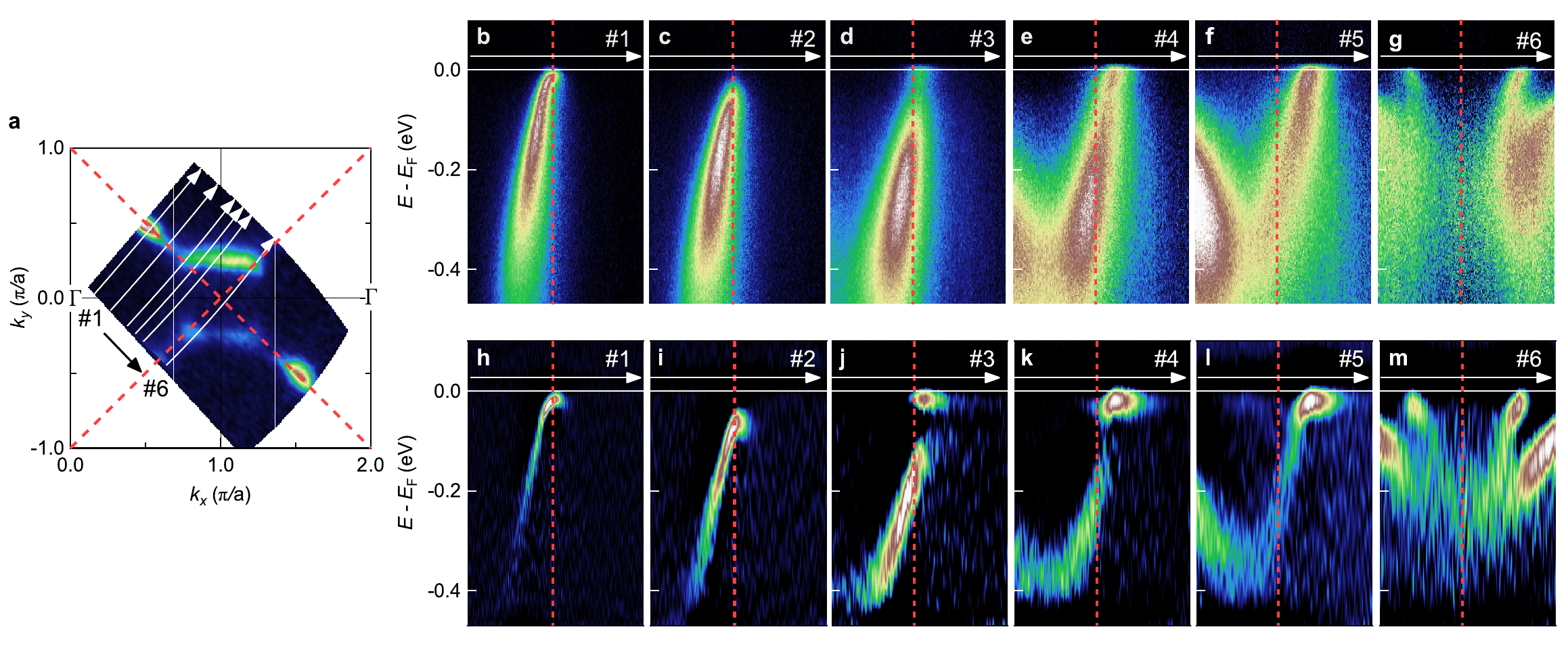}
\end{center}
\caption{\textbf{ARPES spectra of a protect-annealed PLCCO (\boldmath{$x = 0.05$}) single crystal.} \textbf{a}, Fermi surface mapping. Spectral intensity integrated within $\pm 20$ meV of $E_\mathrm{F}$ is plotted. \textbf{b--g}, Band images taken along the cuts indicated in \textbf{a}. The AFM BZ boundary is shown by dashed red lines in each panel. \textbf{h--m}, Corresponding second derivatives with respect to energy.}
\label{PLCCO005}
\end{figure}

\begin{figure}[h]
\begin{center}
\includegraphics[width=0.8\textwidth]{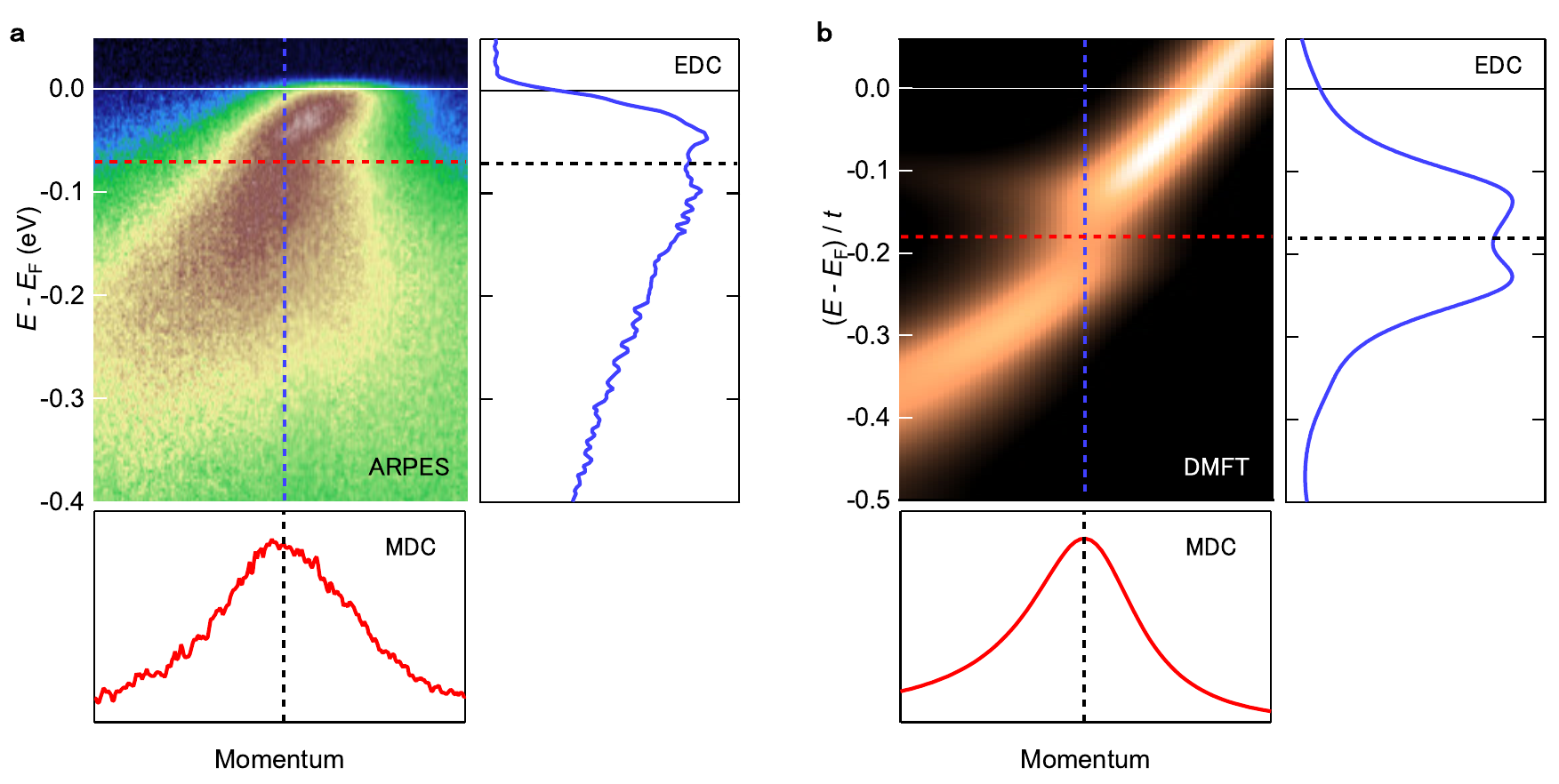}
\end{center}
\caption{\textbf{Determination of the gap positions around ($\pi$, 0).} \textbf{a}, Antinodal ARPES spectra with energy and momentum distribution curves (EDC and MDC, respectively) through the gap point. To define the gap position, one needs to identify the momentum point where the gap is minimized. In between two gapped bands, MDC should be peaked where the magnitude of the gap becomes the smallest. Therefore, the energy-momentum point which falls in the middle of two EDC peaks and coincides with the MDC peak was searched for. By tracking both the EDC and MDC peaks, such a point was identified. \textbf{b}, Procedure for the gap identification applied to the DMFT-calculated results.}
\label{GapPos}
\end{figure}

\begin{figure}[h]
\begin{center}
\includegraphics[width=0.9\textwidth]{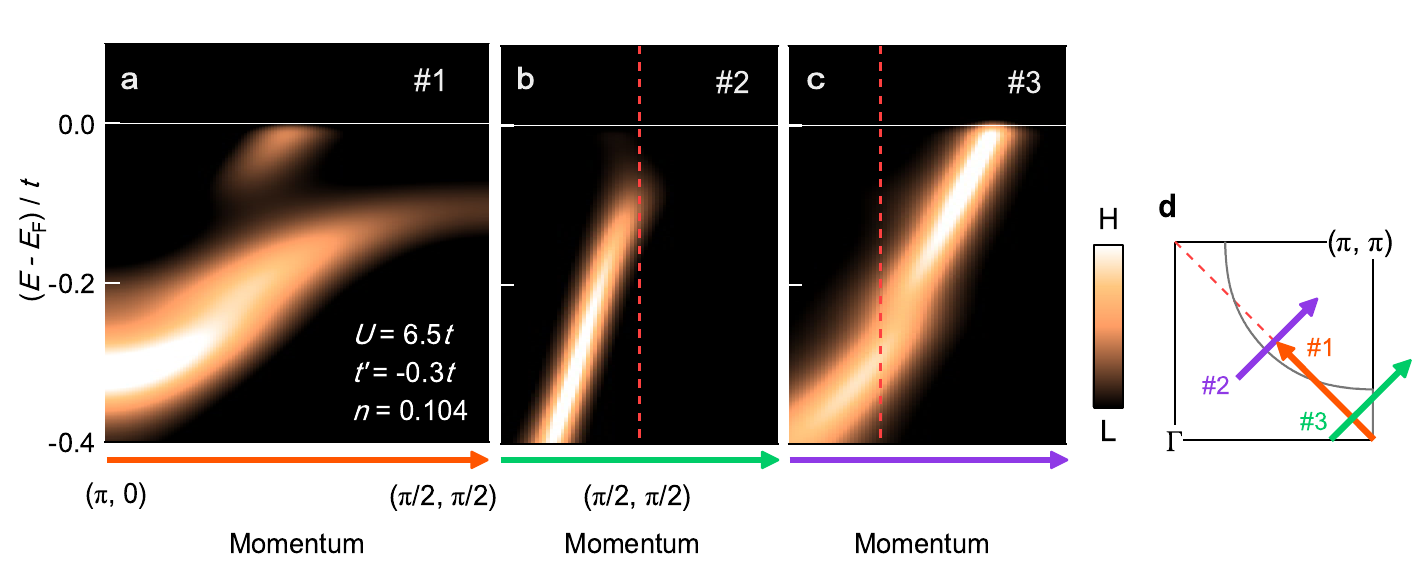}
\end{center}
\caption{\textbf{CDMFT calculations conducted with $U = 6.5t$.} \textbf{a--c}, Calculated spectral function along the cuts \#1--\#3, respectively, indicated in \textbf{d}.}
\label{U65t}
\end{figure}

\begin{figure}[h]
\begin{center}
\includegraphics[width=0.99\textwidth]{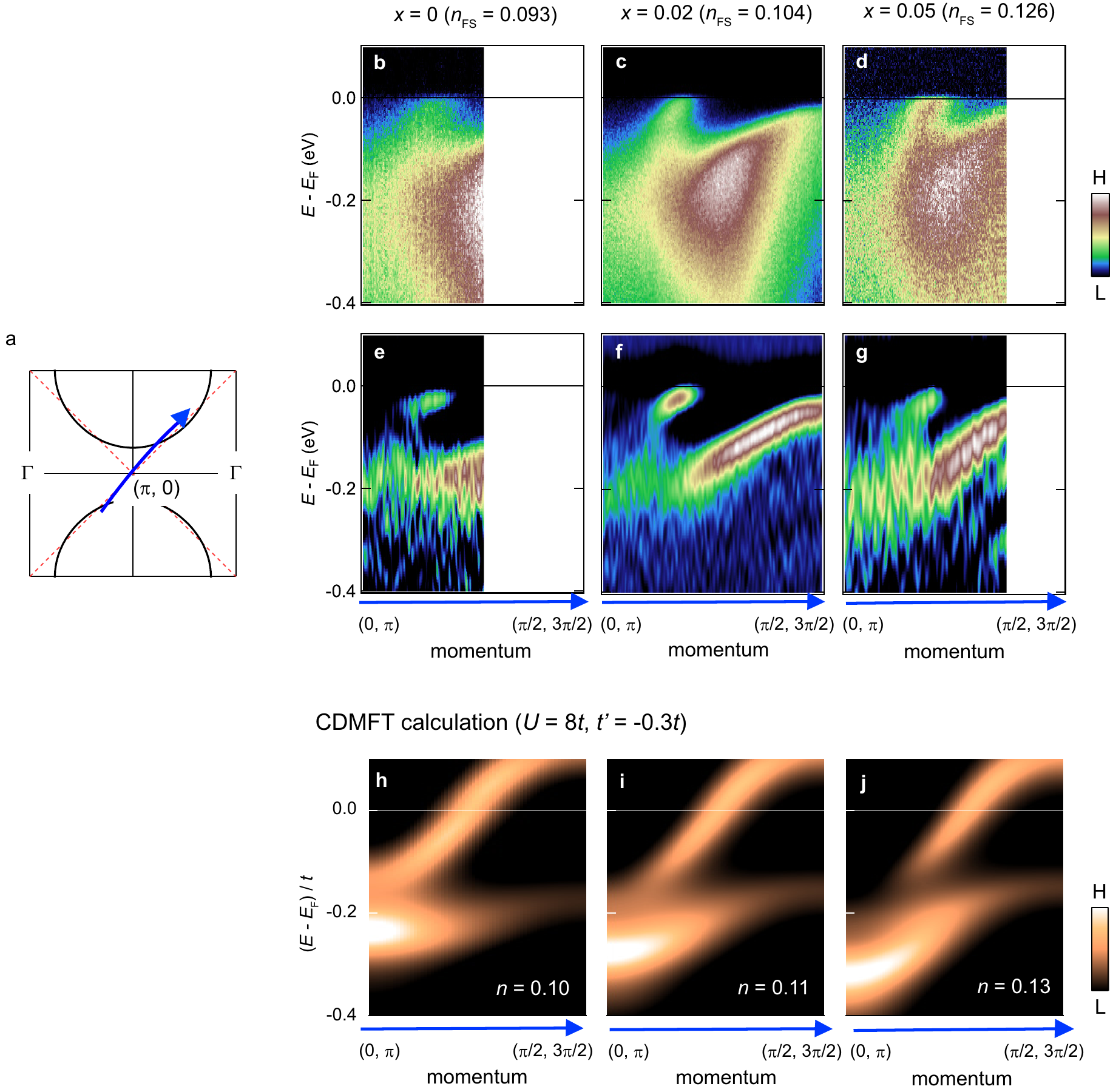}
\end{center}
\caption{\textbf{Band dispersion of protect-annealed PLCCO samples along the AF BZ boundary.} \textbf{a}, A momentum cut for \textbf{b--d} and \textbf{e--g}. \textbf{b--d}, Band dispersion along the cut indicated in \textbf{a} for $x = 0$, $x = 0.02$, and $x = 0.05$ protect-annealed samples, respectively. \textbf{e--g}, Second derivative of \textbf{b--d} with respect to energy. \textbf{h-j}, Spectral function calculated using CDMFT with $U = 8t$ and $t' = -0.3t$. Doped electron concentration $n$ is 0.10, 0.11, and 0.13, respectively.}
\label{DMFT}
\end{figure}

\begin{figure}[h]
\begin{center}
\includegraphics[width=0.99\textwidth]{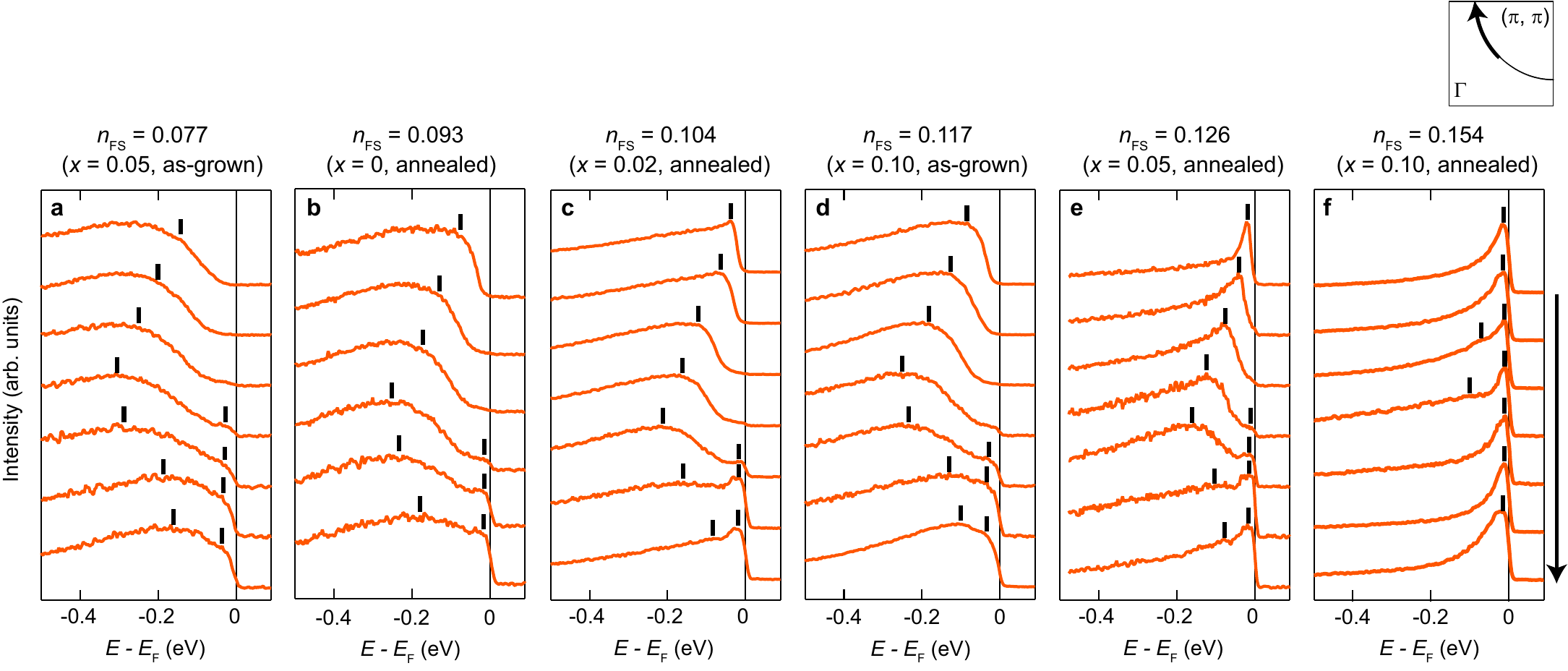}
\end{center}
\caption{\textbf{EDCs of PLCCO samples.} \textbf{a--f}, EDCs plotted from nodal to antinodal region for PLCCO samples with different electron concentration $n_{FS}$ estimated from Fermi surface area. Vertical bars indicates the peak positions.}
\label{EDC}
\end{figure}

\begin{figure}[h]
\begin{center}
\includegraphics[width=0.6\textwidth]{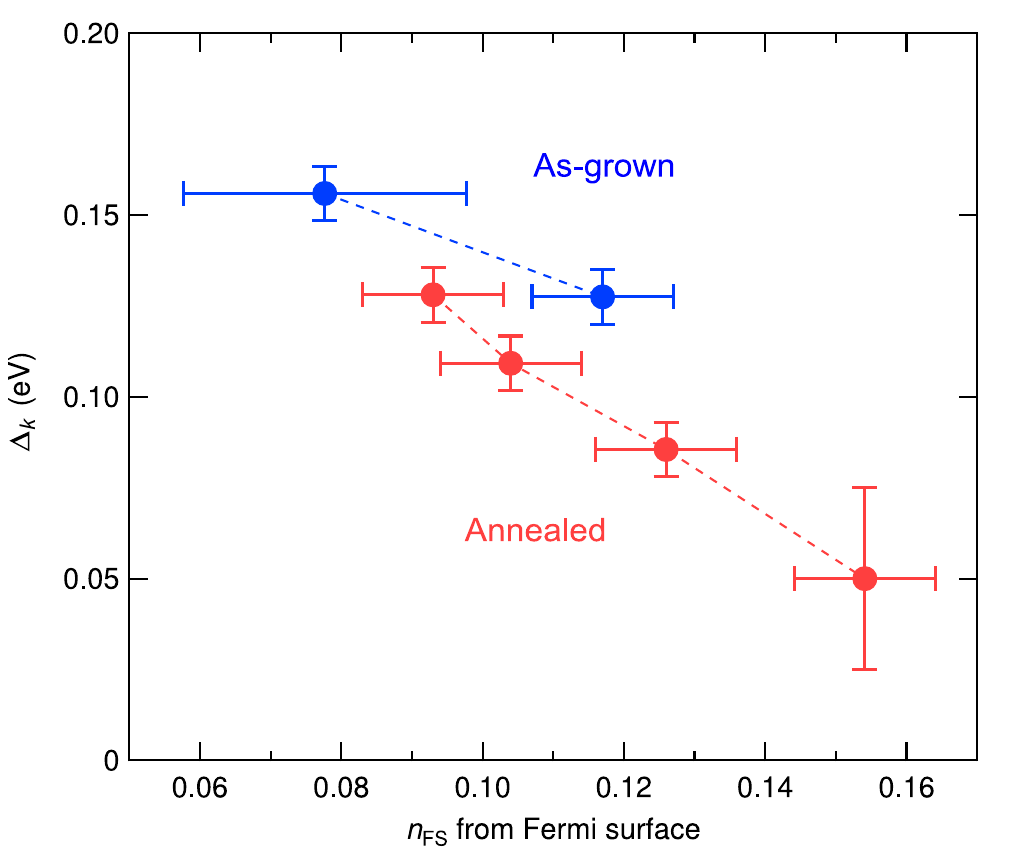}
\end{center}
\caption{\textbf{$\Delta_\mathrm{k}$ versus $n_\mathrm{FS}$ for as-grown and protect-annealed PLCCO samples.} See Supplementary Fig.~\ref{EDC} for the correspondence between $n_\mathrm{FS}$ and the chemical composition.}
\label{DAF}
\end{figure}

\begin{figure}[h]
\begin{center}
\includegraphics[width=0.4\textwidth]{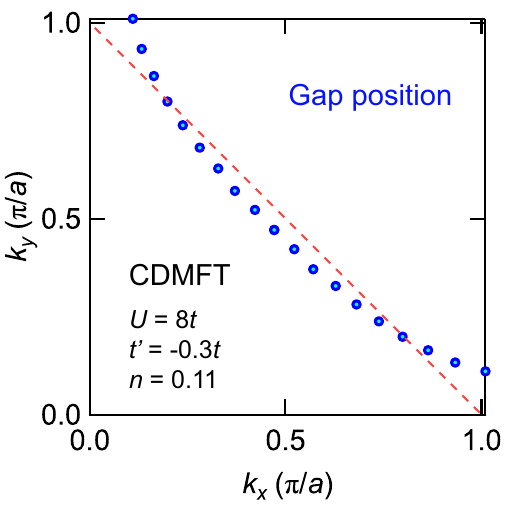}
\end{center}
\caption{\textbf{Gap opening positions for PLCCO (\boldmath{$x = 0.02$}) predicted by the CDMFT calculation.}}
\label{DMFT_gap}
\end{figure}

